\documentclass{aastex631}
\usepackage{natbib}
\usepackage{amsmath}
\usepackage{verbatim}
\usepackage{hyperref}
\usepackage{float}
\usepackage{graphicx}
\usepackage{mathtools}
\usepackage{empheq}
\usepackage{appendix}
\usepackage{microtype}
\usepackage{commath}
\usepackage{bold-extra}
\usepackage{comment}
\usepackage{rotating}
\DeclareGraphicsExtensions{.pdf,.png,.jpg}
\DeclareUnicodeCharacter{2223}{$|$}

\newcommand{\bq}{\begin{equation}} 
\newcommand{\eq}{\end{equation}} 
 
\newcommand{\beq}{\begin{equation}}
\newcommand{\eeq}{\end{equation}}
\newcommand{\beqa}{\begin{eqnarray}}
\newcommand{\eeqa}{\end{eqnarray}}

\newcommand{\PL}{$P$--$L$\ }
\newcommand{\PLs}{$P$--$L$}

\defcitealias{Riess:2023}{R23}
\defcitealias{Riess:2022}{R22}

\shorttitle{Context for Comparing {\it JWST} and {\it HST}}
\shortauthors{Riess et al.}

\begin{document} 

\title{The Perfect Host: {\it JWST} Cepheid Observations in a Background-Free SN Ia Host\\ Confirm No Bias in Hubble-Constant Measurements}

\author[0000-0002-6124-1196]{Adam G.~Riess}
\affiliation{Space Telescope Science Institute, Baltimore, MD 21218, USA}
\affiliation{Department of Physics and Astronomy, Johns Hopkins University, Baltimore, MD 21218, USA}

\author[0000-0002-8623-1082]{Siyang Li}
\affiliation{Department of Physics and Astronomy, Johns Hopkins University, Baltimore, MD 21218, USA}

\author[0000-0002-5259-2314]{Gagandeep S. Anand}
\affiliation{Space Telescope Science Institute, Baltimore, MD 21218, USA}

\author[0000-0001-9420-6525]{Wenlong Yuan}
\affiliation{Department of Physics and Astronomy, Johns Hopkins University, Baltimore, MD 21218, USA}

\author[0000-0003-3889-7709]{Louise Breuval}
\affiliation{Space Telescope Science Institute, Baltimore, MD 21218, USA}

\author[0000-0000-0000-0000]{Stefano Casertano}
\affiliation{Space Telescope Science Institute, Baltimore, MD 21218, USA}

\author[0000-0002-1775-4859]{Lucas M.~Macri}
\affiliation{Department of Physics \& Astronomy, College of Sciences, University of Texas Rio Grande Valley, Edinburg, TX 78539, USA}

\author[0000-0002-4934-5849]{Dan Scolnic}
\affiliation{Department of Physics, Duke University, Durham, NC 27708, USA}

\author[0000-0002-8342-3804]{Yukei S. Murakami}
\affiliation{Department of Physics and Astronomy, Johns Hopkins University, Baltimore, MD 21218, USA}

\author[0000-0003-3460-0103]{Alexei V. Filippenko}
\affiliation{Department of Astronomy, University of California, Berkeley, CA 94720-3411, USA}

\author[0000-0001-5955-2502]{Thomas G. Brink}
\affiliation{Department of Astronomy, University of California, Berkeley, CA 94720-3411, USA}

\begin{abstract}
\vspace{-12pt}
\ \par

Cycle 1 {\it JWST} observations of Cepheids in SN Ia hosts resolved their red-giant-dominated NIR backgrounds, sharply reducing crowding and showing that photometric bias in lower-resolution {\it HST} data does not account for the Hubble tension.
  We present Cycle 2 {\it JWST} observations of $> 100$ Cepheids in NGC$\,$3447, a unique system that pushes this test to the limit by transitioning from low to no background contamination. NGC$\,$3447, an SN Ia host at $D \sim$ 25 Mpc, is an interacting pair comprising (i) a spiral with mixed stellar populations, typical of $H_0$ calibrators, and (ii) a young, star-forming companion (NGC 3447A) devoid of old stars and hence stellar crowding—a rare ``perfect host” for testing photometric bias.
    We detect $\sim 60$ long-period Cepheids in each, enabling a ``three-way comparison'' across {\it HST}, {\it JWST}, and background-free conditions. We find no component-to-component offset ($\sigma < 0.03$ mag; a calibration independent test), and a 50\% reduction in scatter to $\sim$ 0.12 mag in the background-free case, the tightest seen for any SN Ia host.  Across Cycles 1-2 we also measure Cepheids in all SH0ES hosts observed by {\it JWST} (19 hosts of 24 SN Ia; $>$ 50\% of the sample) and find no evidence of bias relative to {\it HST} photometry, including for the most crowded, distant hosts. 
    These observations constitute the most rigorous test yet of Cepheid distances and provide strong evidence for their reliability.
 Combining {\it JWST} Cepheid measurements in 19 hosts (24 SNe~Ia) with {\it HST} data (37 hosts, 42 SNe Ia) yields H$_0  = 73.49 \pm 0.93$ km~s$^{-1}$~Mpc$^{-1}$. Including 35 TRGB-based calibrations (from {\it HST} and {\it JWST}) totals 55 SNe Ia and gives H$_0  = 73.18 \pm 0.88$ km~s$^{-1}$~Mpc$^{-1}$ -- $\sim 6\sigma$ above the $\Lambda$CDM+CMB expectation.



\end{abstract}
\vspace*{-24pt}
\section{Introduction}
\label{sec:intro}

The ``Hubble tension,'' a decade-long discrepancy now reaching a $>5\sigma$ difference between the local determination of H$_0$ and the prediction from $\Lambda$CDM calibrated by the cosmic microwave background (CMB) and which may augur new physics, has motivated additional tests and cross-checks \citep[for extensive and recent reviews of tests, checks, and independent measures, see][]{DiValentino:2021, Tully:2023, Verde:2023, Kamionkowski:2023}. The new capabilities of the {\it James Webb Space Telescope (JWST)} offer the means for additional cross-checks by comparing distances measured to supernova (SN) host galaxies with those measured with the {\it Hubble Space Telescope (HST)}. {\it JWST} has certain distinct advantages compared with {\it HST} for measuring distances to nearby galaxies. For Cepheids, {\it JWST} offers a factor of $\sim 2.5$ higher near-infrared (NIR) resolution than {\it HST} to mitigate crowding from ubiquitous red giants in overlapping old populations, though the optical capability of {\it HST} is still required to find these variables. 

The observed reduction in the scatter of the Cepheid period-luminosity ($P$--$L$) relations based on observations with {\it JWST} is nothing short of remarkable, especially if images are collected across multiple epochs to measure light-curve phases. The enhanced resolution of {\it JWST}  improves the separation between the contaminating background of red giants and Cepheids. In {\it JWST}  Cycle 1, \cite{Riess:2024} found excellent agreement between the $P$--$L$ relations of $> 1000$ Cepheids measured with {\it JWST} and {\it HST} for 5 hosts of 8 SNe~Ia, with both sets measured at the same wavelength, equivalent to NIR $H$-band, with the same slope, and with the same geometric anchor (NGC$\,$4258) yielding a mean difference of $-0.011\pm0.032$~mag. Observations with {\it JWST} in Cycle 2 include observations of the most crowded hosts.  Our Cycle 2 observations also include a unique host that extend the test of Cepheid crowding to its limit, advancing from a regime of low background contamination to none at all.

\begin{figure}[b!] 
\begin{center}
\includegraphics[width=0.85\textwidth]{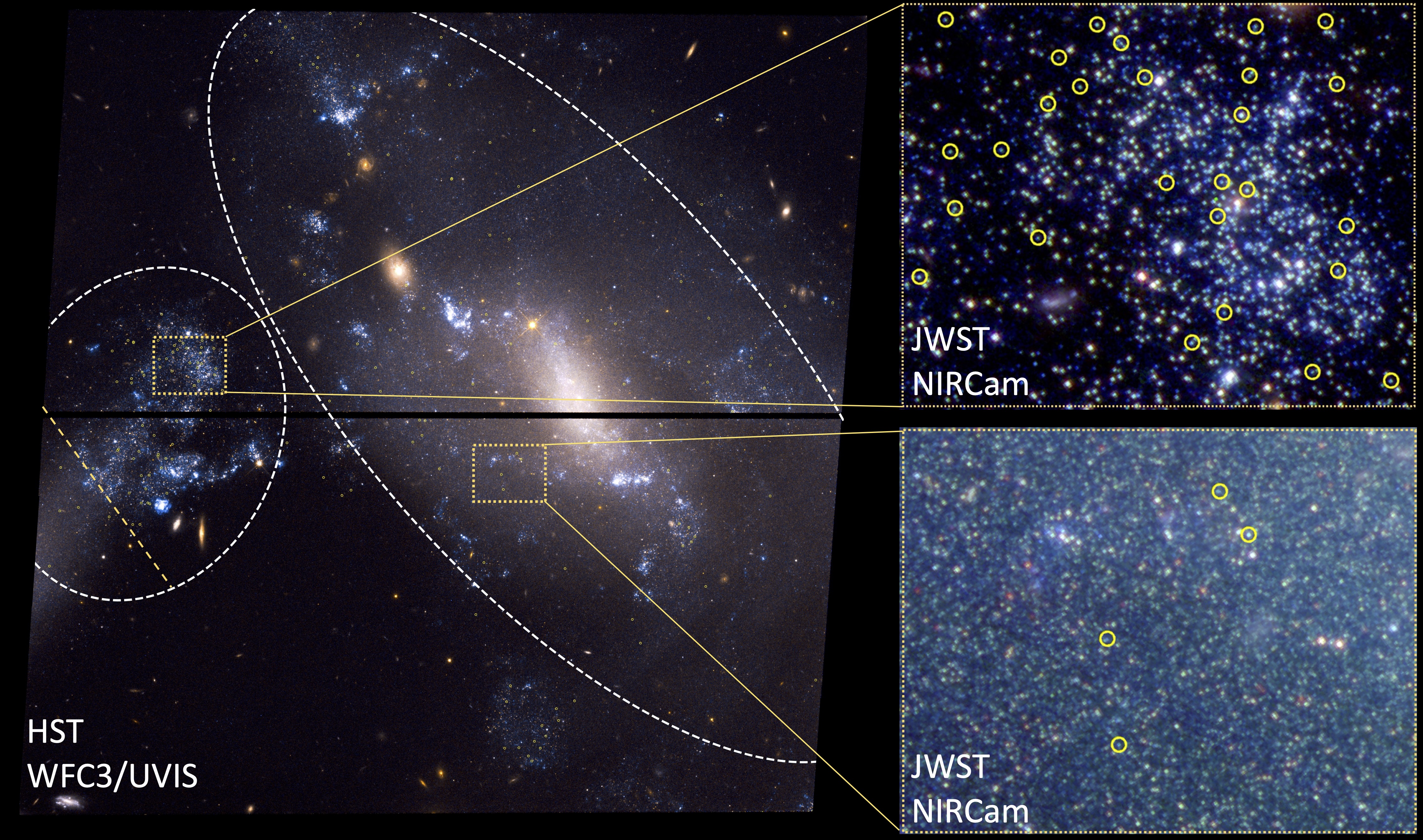}
\end{center}
\caption{\textit{Left:} Color image (red-green-blue) of the NGC$\,$3447 field, generated from {\it HST} WFC3/UVIS imaging ($F814W/F555W+F814W/F555W$). The tidal companion (NGC$\,$3447A; left) and the main body (NGC$\,$3447; right) regions used in this paper are delineated using dashed ellipses (the region to the lower-left of the yellow dashed line in the dwarf region was further excluded, to remove a separate dwarf in projection). \textit{Right:} Zoom-in views that show color {\it JWST}/NIRCam ($F277W/F150W/F090W$) imaging from a region within the dwarf (top) and the main body (bottom). These two cutouts are shown with identical stretch and contrast (drawn from the same parent image) to highlight the lack of crowded stellar background in the dwarf region. Cepheids are identified using yellow circles.}
\label{fg:compositen3447} 
\end{figure} 

\subsection{NGC$\,$3447A as the ``Perfect Host''}

NGC$\,$3447, the host of SN$\,$Ia$\,$2012ht, is one of 37 hosts of 42 Type Ia supernovae (SNe~Ia) in the sample used by \citet[][hereafter R22]{Riess:2022} to measure the Hubble constant and is at the mean distance of the sample, $D \approx 25$ Mpc. It is known to be an interacting galaxy system and in the literature is frequently decomposed into NGC$\,$3447, a disturbed intermediate-luminosity barred spiral, and NGC$\,$3447A, 1.2 mag fainter in the $B$ band, a tidal dwarf-like companion connected by a bridge-like filament as shown in Figure \ref{fg:compositen3447}. Near NGC$\,$3447A, there appears to be another distinct object, with its center lying outside the WFC3 field of view. This third object appears to have a largely older stellar population, but is clearly offset from NGC$\,$3447A in position. We posit that this is a distinct dwarf galaxy overlapping in projection, likely still part of the same galaxy group, but not directly connected to the tidal dwarf NGC$\,$3447A.

According to the study of the NGC$\,$3447 system by \cite{Mazzei:2018}, ``[t]he global morphology suggests, as a working hypothesis, an ongoing head-on encounter.'' Their simulations indicate that both components are part of the same perturbed halo which likely suffered a major encounter with another massive halo $\sim 300$ kpc away, recently producing gravitational instabilities in the NGC$\,$3447/3447A and tidal star formation, with a $B$-band-weighted stellar population age of 1.3 Gyr. They also provide a list of galaxies residing in the same group with perturbed morphologies, indicating there is no lack of candidates for the encounter with NGC$\,$3447. 

While the origin story for the system presented by \cite{Mazzei:2018} appears plausible, we note that its accuracy is not of relevance for the utility of this system for testing Cepheid distances. Regardless of the true origin of the system, the key point relevant to this paper is that the object we refer to as NGC$\,$3447A is a distant ($D > 20$ Mpc) SN Ia host containing an abundant collection of Cepheid variables which are not superimposed on an older stellar population, thus effectively eliminating the primary source of crowding in the {\it JWST} imaging. Indeed, such systems, while rare, are not that unusual, such as tidal dwarf galaxies (TDGs) in interacting systems \citep[e.g., NGC$\,$5291, NGC$\,$4694, NGC$\,$7252;][]{Lelli:2015}, which consist of young stars formed {\it in situ} from the tidally expelled gas. It is this lack of the old, red-giant background that allows us to provide a stringent test on the effect of crowding on the photometry of Cepheid variables in SN~Ia hosts.

In \S2 we present the data, primarily stellar photometry and metallicity measurements in the NGC$\,$3447 system. \S3 contains an analysis of the stellar populations of the components, the Cepheid $P$--$L$ relations, and findings regarding their relative distance measures. We also provide comparative results for all JWST Cepheid/SN Ia hosts observed since Cycle 2.  We present a discussion in \S4.

\section{Observations}

We obtained observations of NGC$\,$3447 with NIRCAM on {\it JWST} as part of GO program 2875 \citep{2023jwst.prop.2875R} on 2024 May 23 (14--15 hr) UTC in 4 filters: $F090W/F356W$ for 1890 s and $F150W/F277W$ for 2534 s. Each image was taken at 4 dither positions, with the B detector centered on the host and the A detector well off the host.
The data were calibrated with Data Build 10 (though there are no changes significant for our data in Build 11).  

\subsection{Photometry}
Photometry was measured with the point-spread-function (PSF)-fitting package DOLPHOT and its NIRCam module \citep{2016ascl.soft08013D,2024ApJS..271...47W} using the methodology and steps given by \cite{Riess:2023} (hereafter R23) and \cite{Riess:2024}. 

We adopt elliptical contours to refer to sources in the region of the two main components of NGC$\,$3447 following \cite{Mazzei:2018}. As we will show, our findings are not sensitive to the exact specifications of the two regions; differences only have the potential to change the sample by a few percent. For NGC$\,$3447 we use a center at $\alpha=163.350^\circ, \delta=16.774^\circ$ (J2000), position angle (PA) $=95^\circ$, major axis $=1.6\arcmin$, minor axis $=0.7\arcmin$. For NGC$\,$3447A  we use a center at $\alpha=163.372^\circ, \delta=16.786^\circ$, PA $=-30^\circ$, major axis $=0.6\arcmin$, minor axis $=0.5\arcmin$. We add a chord to the definition of NGC$\,$3447A to section off contributions to the edge of NGC$\,$3447A from the old diffuse dwarf seen in projection as shown in Figures \ref{fg:compositen3447} and \ref{fg:metallicity_map} and discussed in \S1.1. 

\subsection{Metallicity}

A relatively flat metallicity gradient between 0 and 10 kpc from the center was measured from H II regions in the NGC$\,$3447 system by the SH0ES Team as presented by \cite{Hoffmann:2016}. Following the discovery of the importance of the NGC$\,$3447A system, an additional set of metallicity measurements was added on 2025 Feb. 27 UTC with the Low Resolution Imaging Spectrometer (LRIS; \cite{Oke:1995}) on the Keck-I 10\,m telescope. These metallicity measurements were made using the method described by R22..

\begin{figure}[b!] 
\begin{center}
\vspace*{12pt}
\includegraphics[width=\textwidth]{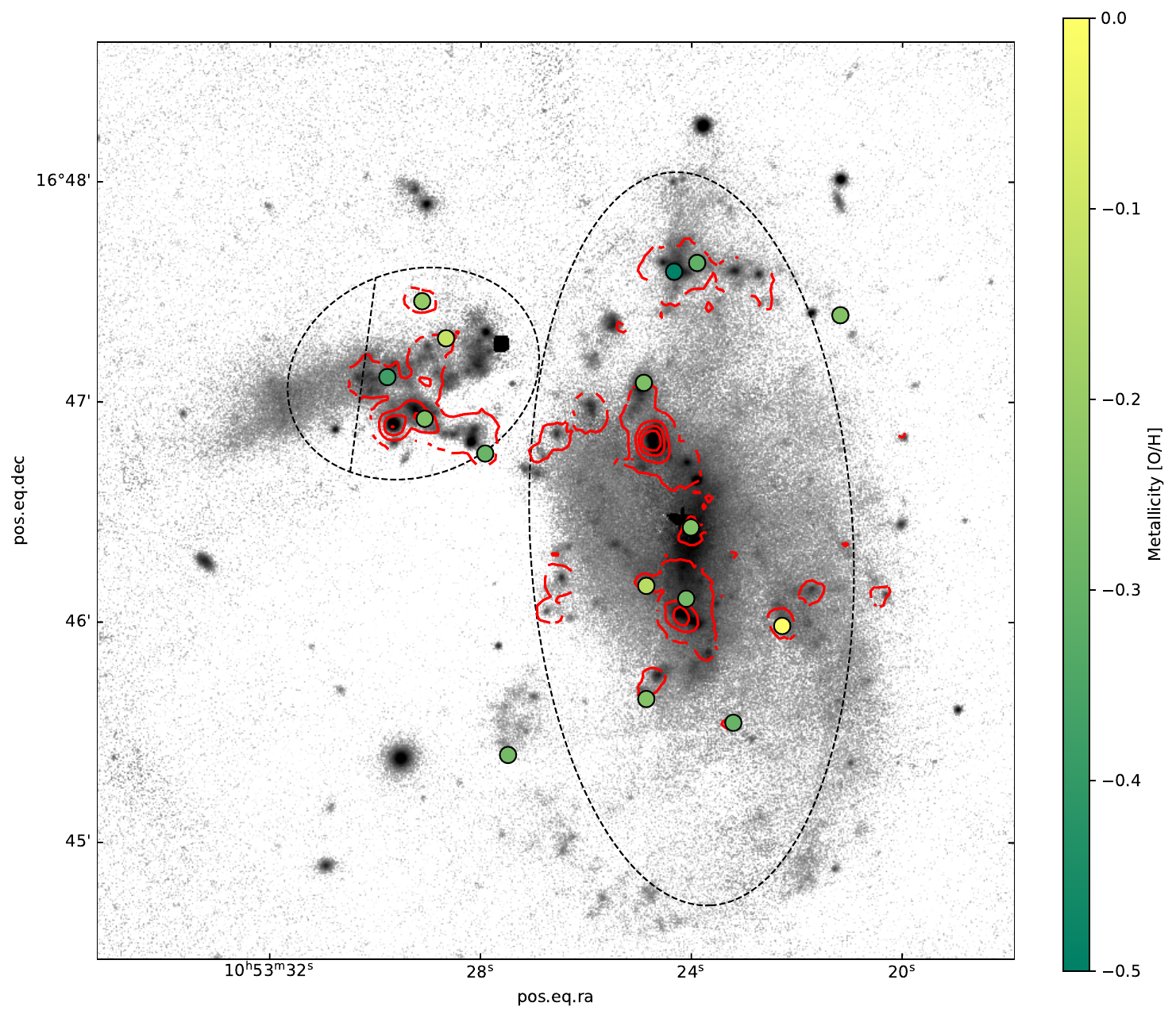}
\end{center}
\caption{Metallicity map and H$\alpha$ image \citep[from][]{Mazzei:2018}.  The metallicity measures were obtained using the method described by R22 (see also \citealt{Hoffmann:2016}) from strong nebular lines in H~II regions at the locations indicated using Keck/LRIS.}
\label{fg:metallicity_map} 
\end{figure}

The full set of 16 measurements is shown in Figure \ref{fg:metallicity_map}. The augmented map includes 5 separate measures in the tidal host. As shown, there is little apparent difference between the spiral and the tidal companion, and both have a sample mean of [O/H] = $-0.25$ dex. This similarity is an important diagnostic for the origin of the NGC$\,$3447A system. If NGC$\,$3447A were an  independent, primordial dwarf system on its first encounter with NGC$\,$3447 it would have a much lower metallicity than NGC$\,$3447. Rather, the tidal star formation of NGC$\,$3447A is largely from the same reservoir of gas and dust, and the enrichment likely occurred early in the gaseous disk of the parent galaxy before the interaction that led to the existence of NGC$\,$3447A, which would inherit a similar metallicity \citep{Lelli:2015}. Regardless of the source of the enrichment and origin of NGC$\,$3447A, the similar measured metallicity for the two components means we can directly compare the Cepheid magnitudes in each without regard to the Cepheid metallicity relation \citep{Breuval:2024, Breuval:2025}.

\section{Analysis}

\subsection{Stellar Populations in the NGC$\,$3447 System}

\begin{figure}[t!] 
\begin{center}
\includegraphics[width=\textwidth]{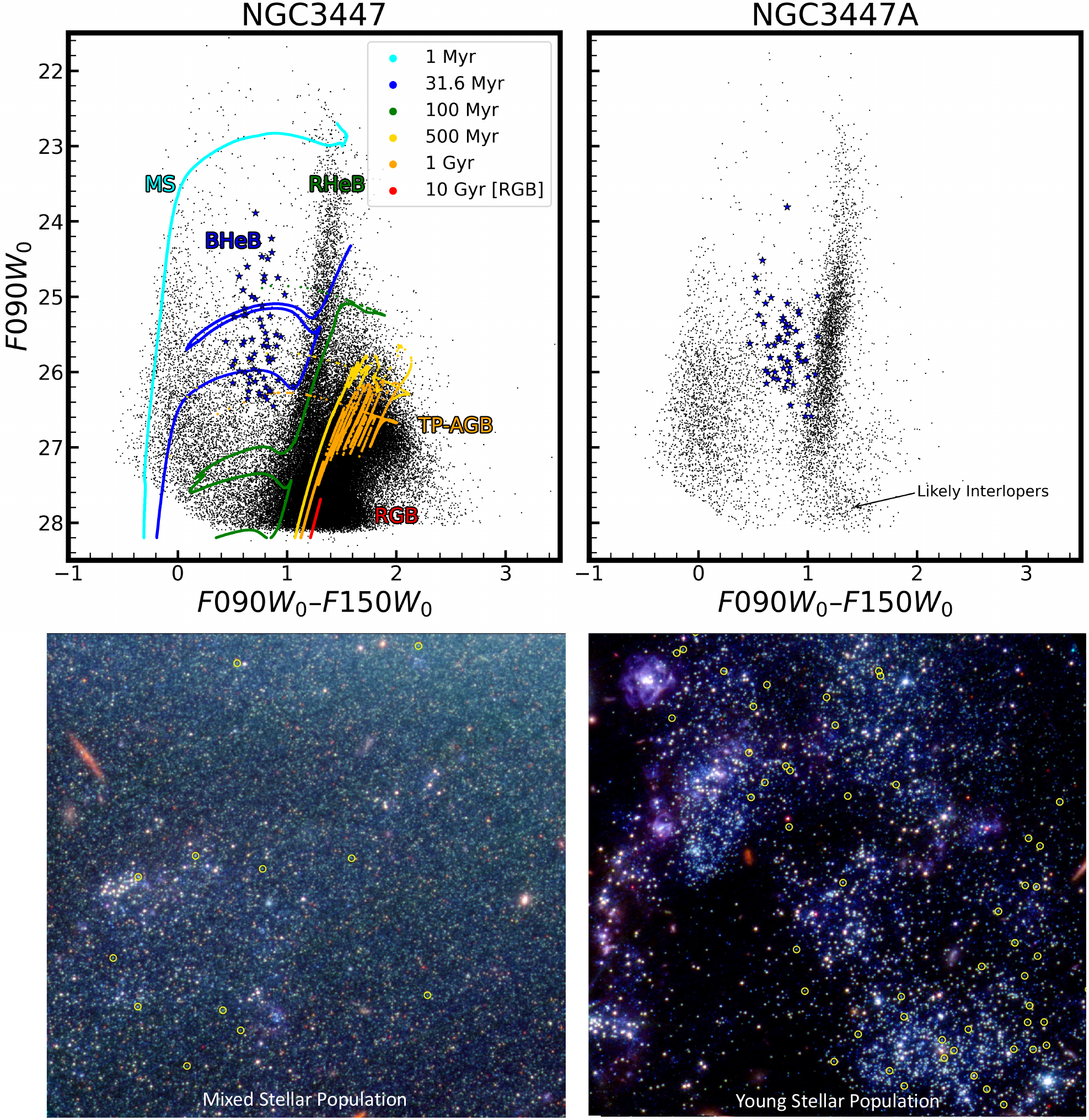}
\end{center}
\caption{\textit{Top left:} CMD of the main spiral body (NGC$\,$3447). Overlaid are example stellar sequences drawn using the ArtPop software package \citep{2022ApJ...941...26G}. The main spiral body shows clear signs of a broad age of stellar populations, from young main-sequence stars to old red giants. \textit{Top right}: CMD of the selected tidal dwarf region. Unlike the main body, the stellar population here is dominated by young and intermediate-age stars ($\sim 100$ Myr). The few red giants are likely interlopers from the halo of the main spiral and the separate dwarf found in projection to the system. \textit{Bottom panels:} Example color-image cutouts drawn from the main body and the tidal dwarf, showing the clear visible differences in stellar populations -- the tidal dwarf region lacks the background, old stellar population that can be seen in the disk of the main galaxy.}
\label{fg:stellar-pops} 
\vspace*{12pt}
\end{figure}

Our {\it JWST} photometry allows us to examine the stellar populations throughout the NGC$\,$3447 system which may inform our understanding of the two Cepheid regions we compare in the section that follows. Figure \ref{fg:stellar-pops} shows the color-magnitude diagrams (CMDs) for {\it JWST} photometry $F090W$ versus $F090W-F150W$, dereddened for Milky Way-type (MW) reddening based on $E(B-V)=0.026$ mag \citep{Schlafly:2011} and the reddening relations for MW-type reddening \citep{Anand:2024}, 
\begin{equation} \Delta A_{F090W} = 1.416\ \Delta E(B-V)\end{equation}
\begin{equation} \Delta A_{F150W} = 0.602\ \Delta E(B-V). \end{equation}

 On the first CMD (corresponding to the main body, NGC$\,$3447; left panel), we show stellar populations of varying ages drawn using MIST isochrones \citep{2016ApJS..222....8D} from the ArtPop software package \citep{2022ApJ...941...26G}. This main body of the spiral clearly shows stars ranging from very young, bright, and blue main-sequence stars ($\sim 1$~Myr), to red, fainter, and much older red-giant-branch stars ($\sim 10$~Gyr). This CMD looks similar to those of other SH0ES spiral hosts as seen in \cite{Riess:2024} and thus represents typical scenes containing Cepheids (though as we show in the next section, a little more diffuse than some).

\begin{figure}[b!] 
\begin{center}
\includegraphics[width=0.9\textwidth]{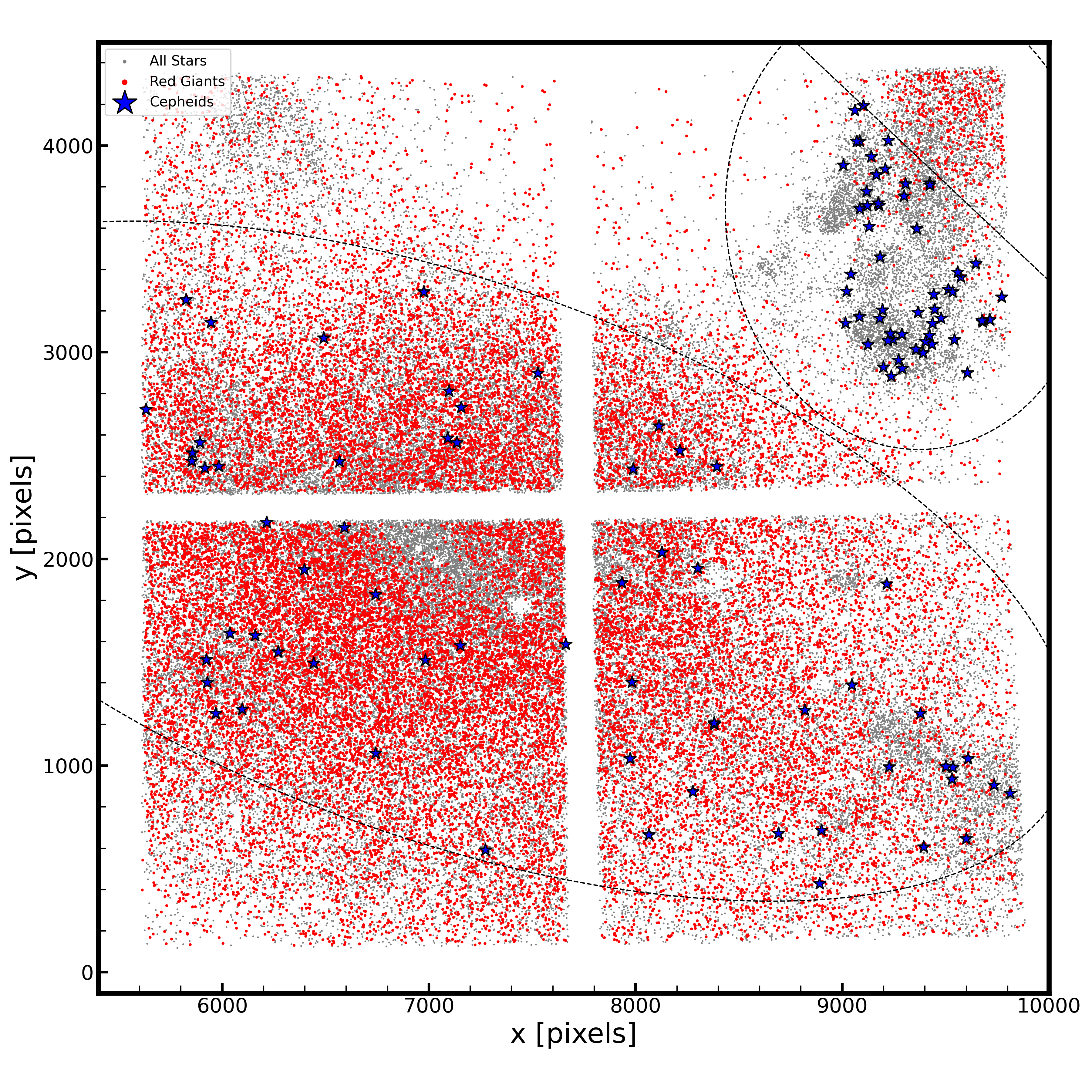}
\end{center}
\caption{Color-coded identity of stars in the NIRCAM field of NGC$\,$3447. Regions follow the definitions in \S2.1.  As shown, the Cepheids in tidal host NGC$\,$3447A (upper right) are free of contamination from common RGB stars with $F090W_0 > 27.55$ mag and $0.7< $ (F090W-F150W)$_0$ $<2.0$ mag. }
\label{fg:starsbycolor} 
\end{figure}

In contrast, the tidal companion NGC$\,$3447A shows a strictly young stellar population (up to $\sim 100$~Myr in age), with a distinct paucity of intermediate-age asymptotic giant branch (AGB) and old red giant branch (RGB) stars. Figure \ref{fg:starsbycolor} displays the location of Cepheids and RGB stars throughout the field of view of the primary NIRCam module covering this system. The few RGB stars visible in the tidal companion's CMD are most likely interloping stars from both the outer halo of the main spiral and the separate, older dwarf likely in projection (as seen in Figure \ref{fg:compositen3447}). It is evident that NGC$\,$3447A has only recently formed stars (and quite intensely based on the high Cepheid density) owing to the tidal interaction. It is this lack of intermediate- and old-aged stars in the companion NGC$\,$3447A that eliminates the apparent crowding of Cepheid variables, making it the ``perfect host'' for measuring Cepheids.

The embedded red super-giant branch seen in both galaxies shows significantly greater reddening in the spiral by $\Delta E{(B-V)}=0.07$ mag or $\Delta E{(V-I)}=0.09$ mag, a likely consequence of the highly efficient conversion of gas and dust to stars in the tidal host. Having less dust in NGC$\,$3447A can only improve Cepheid distance measurements, with less dust causing less variation in color-to-dust relations. We will compare the difference in the RGB to the Cepheid colors in the next section.

\subsection{Cepheids}

\begin{figure}[h!] 
\begin{center}
\includegraphics[width=\textwidth]{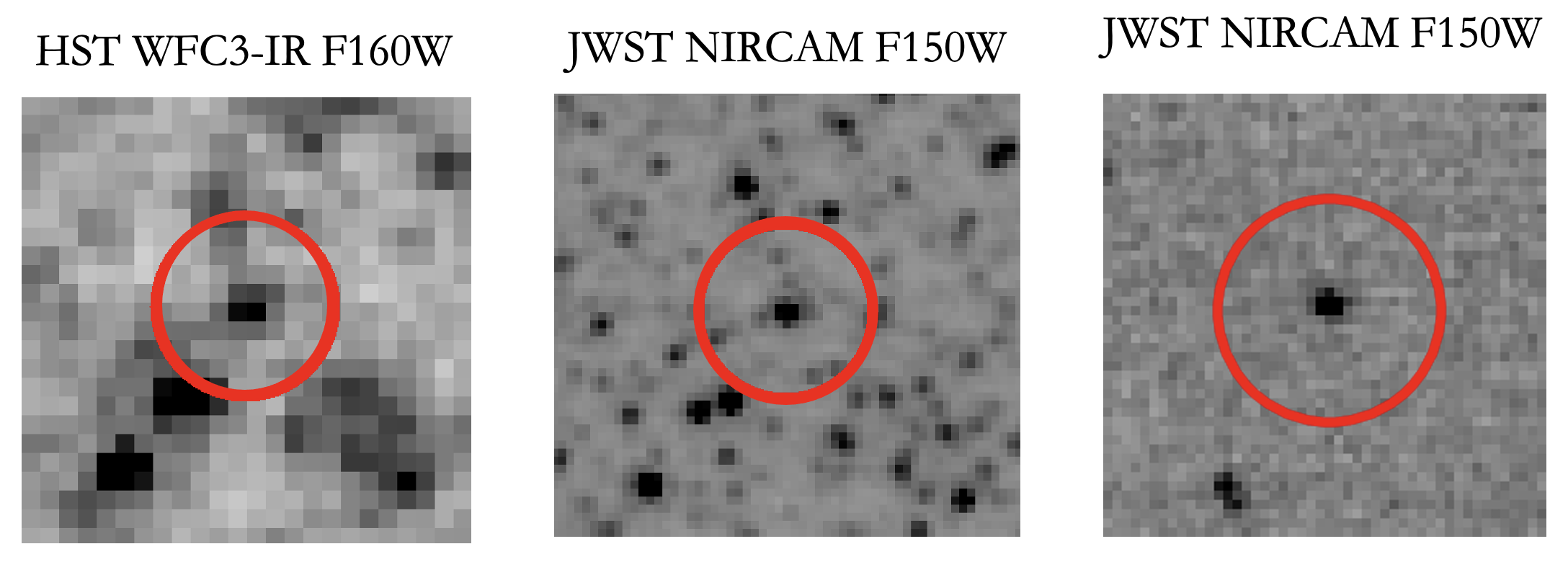}
\end{center}
\caption{NIR postage stamps of Cepheids in NGC$\,$3447 which illustrate the three levels of crowding in this study.  The left panel shows a Cepheid observed with {\it HST} in the spiral component of NGC$\,$3447. In the middle panel, we see the same Cepheid observed with {\it JWST} NIRCAM at the same wavelength. In the right-hand panel is a Cepheid of similar period in the tidal host, NGC$\,$3447A, illustrating the lack of a stellar background.}
\label{fg:stamps} 
\end{figure}

\begin{figure}[b!] 
\begin{center}
\includegraphics[width=\textwidth]{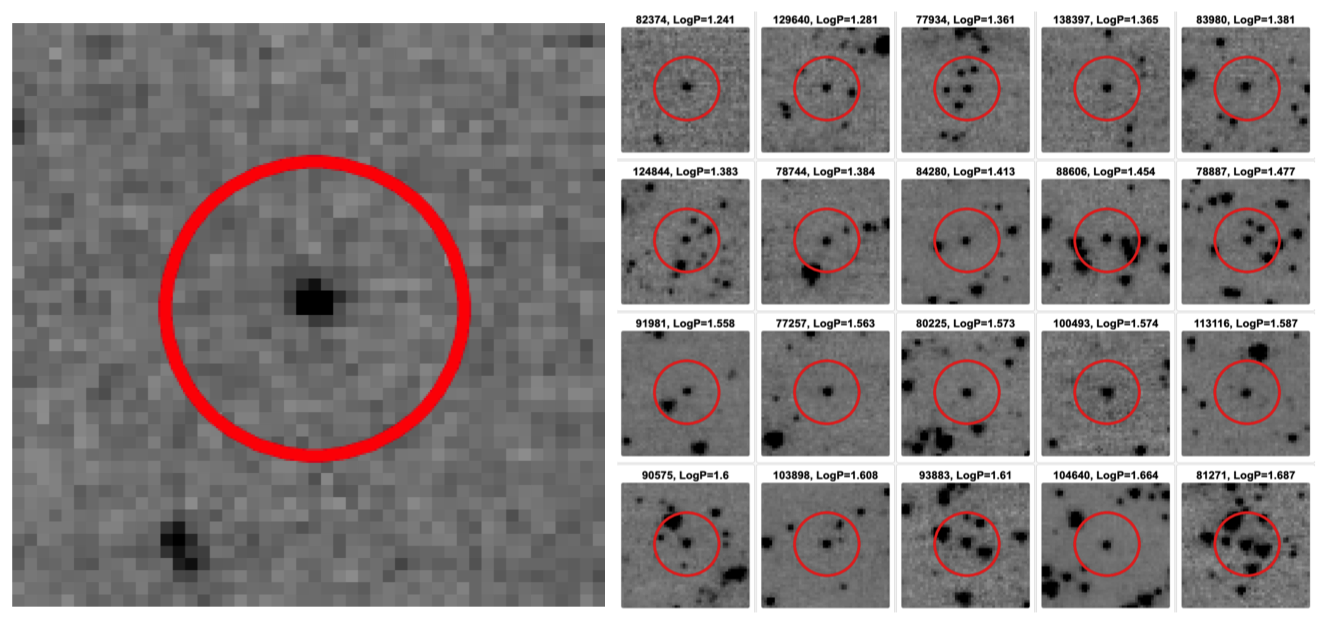}
\end{center}
\caption{A larger sample of Cepheids in NGC$\,$3447A, the tidal companion, illustrating the absence of a contaminating background.  The general appearance is similar to that of a MW open cluster.  As discussed in the text, the rare superposition of a blue star with a Cepheid would suppress the latter's amplitude, more strongly in the optical (where their light curves are measured) than in the infrared, and eliminate such an object from the amplitude selection.}
\label{fg:mostamps} 
\end{figure}

As shown in the right-hand panel of Figure \ref{fg:starsbycolor}, the density contrast of Cepheids to red giants is several orders of magnitude higher in NGC$\,$3447A compared to NGC$\,$3447. The lack of background stars in NGC$\,$3447A, understood from the absence of an old population in this young component, is readily apparent from the image stamps in Figures \ref{fg:stamps} and  \ref{fg:mostamps}. The only other stars in the Cepheid fields in NGC$\,$3447A are blue main-sequence stars or blue and red helium-branch stars, all of which are quite sparse (existing only at the top of the stellar mass function) and thus rarely overlap a Cepheid\footnote{Strong superposition of a bright blue star with a Cepheid would have an even greater impact in the optical ($F555W$) where Cepheids are found, depressing the amplitudes of their light curves and thus eliminating such Cepheids with the rare overlap from the partially-amplitude-selected Cepheid sample.}. Background fluctuations appear at the pixel scale rather than the image scale, dominated by detector read noise (including $1/f$ noise) rather than background sources as seen in the right-hand panel of Figure \ref{fg:stamps}. 

To quantify the sparse, visual impression we perform artificial-star tests as described by R23. In NGC$\,$3447A, we measure the mean crowding bias (input minus output) in $F150W$ to be $\sim 0.005$ mag, substantially lower than seen anywhere beyond the Local Group in the NIR (as far as we are aware). For reference, \cite{Riess:2012} measured a mean crowding bias of 0.002 mag for Cepheids in M31 with {\it HST} at the same wavelength. 

\begin{figure}[b!] 
\begin{center}
\includegraphics[width=\textwidth]{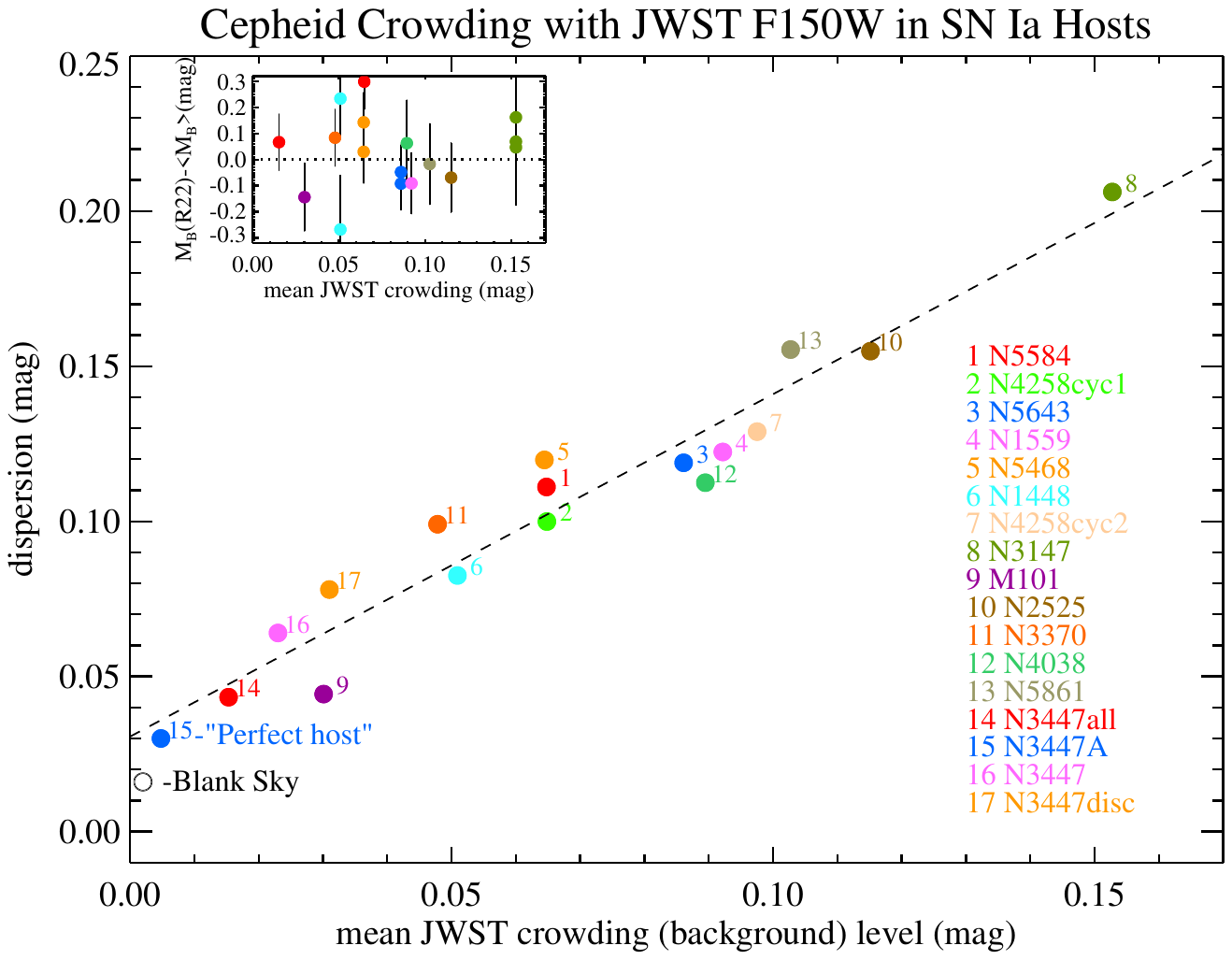}
\end{center}
\caption{Comparison of the mean {\it JWST} background due to crowding (as measured from artificial stars) and the crowding noise (measured from the dispersion of artificial stars). The two quantities are closely related as they are both measures of the stellar background.  Several regions in NGC$\,$3447 are shown including the tidal host (``perfect host'') which is very similar to the ``blank sky'' measured from the opposite (host-free) NIRCAM module.  The inset plots the crowding level versus the residual of each hosted SN from the distance-ladder fit of R22.  We find no correlation between these quantities and thus no evidence for an uncorrected bias in {\it HST} photometry with crowding.}
\label{fg:crowd} 
\end{figure}

The mean dispersion of retrieved stars (``crowding noise'') is also very small, $\sim 0.03$ mag, despite including contributions from the sky and detector read noise. In Figure \ref{fg:crowd}, we also show the results for two other regions in the NGC$\,$3447 system, the area contained in the NGC$\,$3447 ellipse (N3447) and a region around the spiral componen which is $\sim$ 30\% smaller than the previously-defined ellipse (``N3447disc''). To quantify the level of noise from the ``blank sky,'' we also added fake stars to NIRCAM module A which did not contain NGC$\,$3447 and show that point in Figure \ref{fg:crowd} at crowding = 0.002~mag and dispersion = 0.016~mag. As indicated by the imaging,  the background of NGC$\,$3447A is much closer to that of the blank sky than to the next sparsest (and nearest) SN host, M101. 

In Figure \ref{fg:crowd} we show the relation between crowding bias and artificial-star dispersion for all {\it JWST} Cycles 1 and 2 targets observed through $F150W$. As expected, there is a very tight relation between the two as they both are a measure of the stellar background. The median crowding-induced dispersion of $\sim 0.10$~mag is still one of the main contributors to the full, apparent \PL dispersion of $\sim 0.16$--0.18~mag. We estimate additional contributors to \PL dispersion to be the quality of phase correction, intrinsic scatter (from the width of the instability strip), measurement signal-to-noise ratio, and inhomogeneous extinction, in approximate descending order of importance (R23).

\begin{figure}[b!] 
\begin{center}
\includegraphics[width=\textwidth]{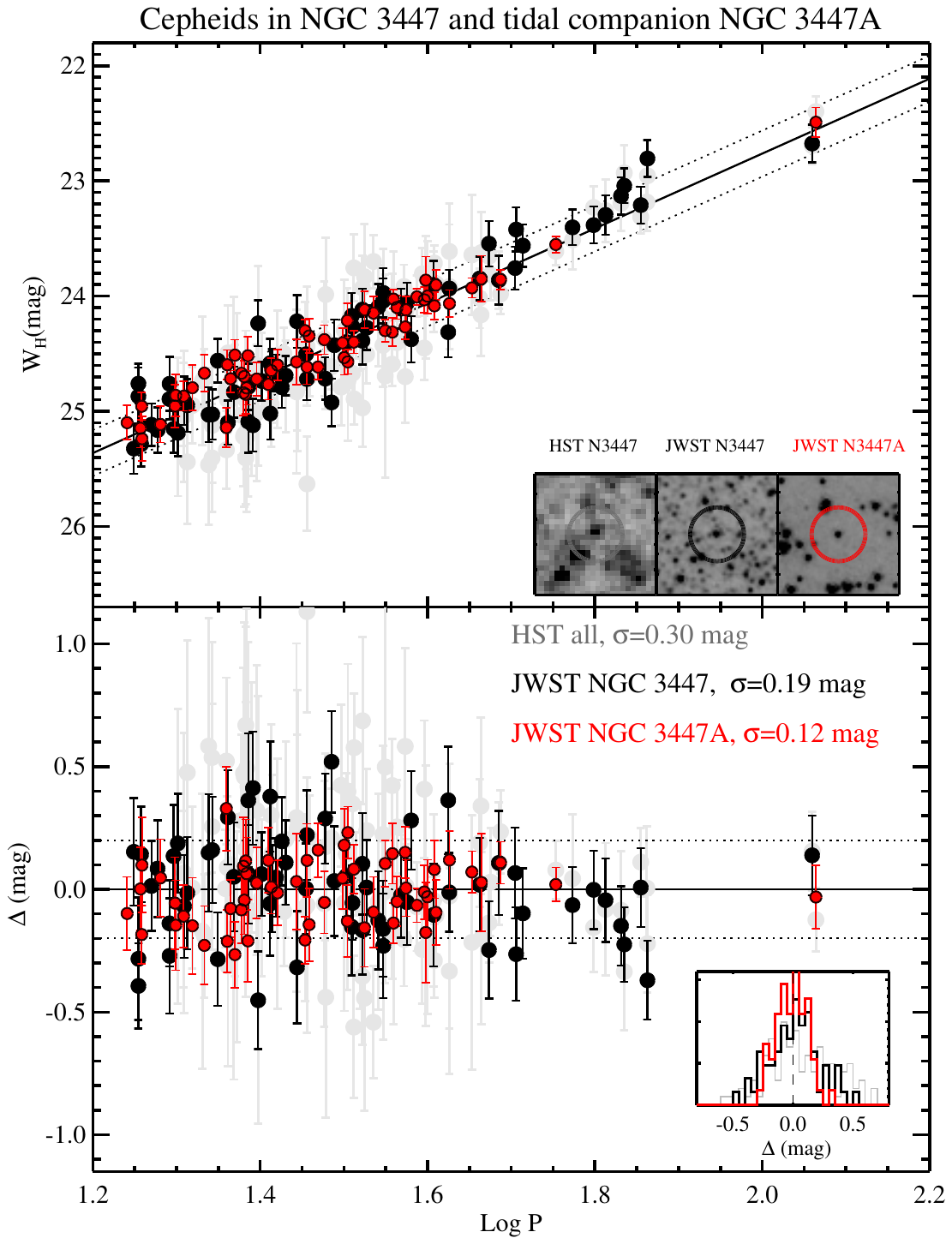}
\end{center}
\vspace*{-12pt}
\caption{Reddening-free (Wesenheit) $P$--$L$ relations for three levels of crowding in NGC$\,$3447.  The points in light gray are from {\it HST} measurements in $F160W$ as given by R22.  In blue are Cepheids observed with {\it JWST} $F150W$ in the spiral host and in red are those observed in the tidal host, NGC$\,$3447A.  All three samples yield consistent distance measures but with decreasing levels of \PL noise.}
\label{fg:n3447pl} 
\end{figure} 

We note that the level of crowding for all {\it JWST} targets from SH0ES is below the size of the Hubble tension ($\sim 0.17$~mag), and the (correlated) uncertainty in the {\it JWST} crowding correction, $\sim 20$\% of the correction (R22), is even lower with a mean per host for {\it JWST} of $\sim 0.02$ mag. All of the {\it JWST} targets (including the most crowded, NGC$\,$3147) are below the level of the {\it least} crowding of any {\it HST} target, which, not surprisingly, was NGC$\,$3447, at 0.17 mag. In the inset to Figure \ref{fg:crowd}, we plot the {\it JWST} crowding versus each SN's $M_B$ relative to the mean ($M_B$ measured with {\it HST} by R22), which shows no correlation between the two (0.6$\sigma$ trend, in the direction of overcorrecting {\it HST} observations for crowding).  That is, there is no evidence for greater or lesser crowding producing a trend with the crowding-corrected distance measurements.

\subsubsection{JWST and a Single-Epoch Phase Correction}
In {\it JWST} Cycle 1 we observed hosts with two epochs separated by a few weeks to enable phase recovery to reduce the \PL dispersion over the use of a random phase (R23). However, multi-epoch observations with {\it JWST} are costly as the telescope has long slew and settling times.

In Cycle 2 we obtained a single epoch and used a different method to constrain the phase of the observation. Because each Cepheid has a previously measured period, mean magnitude, and amplitude in the {\it HST} WFC3-UVIS filter $F814W$, a band which is close in wavelength to NIRCAM $F090W$ (also to $F070W$), we can constrain the random observation phase with {\it JWST} of each Cepheid by determining at which phase the single $F090W$ observed magnitude best matches the individual, known $F814W$ light curves. This requires only a small color transformation (from $F555W$) which we determined synthetically from Cepheid spectral energy distributions (SEDs):
$F090W = 1.18(F814W)-0.18(F555W)$. Thus, comparing the measured single-epoch $F090W$ to the expected $F090W$ light curve identifies the best-matching phase. That phase is then assigned to the primary filter, $F150W$, observed at the same time and which is used in the $W_H$ magnitude.

For NGC$\,$3447, we find the set of 154 individual phase corrections have a mean value (corrected minus uncorrected) of 0.003 mag and a dispersion of 0.070 mag. As expected, the phase correction has no net impact on the distance (i.e., the mean of random phases is the mean phase), but it usefully reduces the \PL dispersion. For $F090W$, where the phase correction is derived, the phase-correction mean is 0.012 with a dispersion of 0.15 mag. Shorter wavelength measurements provide greater leverage for estimating the phase owing to their greater amplitudes. In practice, the single-epoch phase corrections are seen to reduce the \PL dispersion from 0.18 to 0.17 mag in $F150W$, from 0.27 to 0.24 mag in $F090W$, and from 0.19 to 0.17 mag in $W_H$, equivalent in statistical weight of increasing the Cepheid sample size by $\sim 25$\%. The improvement is comparable to the use of two epochs by \cite{Riess:2024} where the mean \PL dispersion of two epochs decreased from 0.19 to 0.18 mag with phase correction. Because phase noise adds to other noise sources, the impact of phase correction on the total dispersion will depend on the relative size of other sources of scatter and would be negligible for high-scatter \PLs.

\subsubsection{Period-Luminosity Relations}

In Figure \ref{fg:n3447pl} we show the Wesenheit magnitude, $W_H=H-0.4(V-I)$, versus log(period) relation, which is the conventional, dereddened SH0ES photometric system relation for  Cepheids for {\it HST} and {\it JWST} as described by R23, where $H=F150W$ for {\it JWST} and $H=F160W$ for {\it HST}, and $V-I=F555W-F814W$. For these samples we employ a 3$\sigma$ rejection, a conservative threshold following Chauvenet's criterion as defined for these sample sizes. 

The dispersion of the {\it HST} Cepheid magnitudes in the NGC$\,$3447 system is 0.30 mag, at the low end for SN~Ia hosts observed by {\it HST}. This fact already indicates the diffuse nature of the system, though this was not previously noted by \cite{Riess:2016} and R22, nor by \cite{Hoffmann:2016}. The {\it HST} NIR observations have some reduction in phase noise because they were averaged across five well-spaced phase points versus one for {\it JWST}. 

For NGC$\,$3447, the spiral host, the $W_H$ \PL dispersion is 0.194 mag, comparable to that seen in other spirals hosts observed with {\it JWST} \citep{Riess:2024}. For NGC$\,$3447A, there is a substantial drop in dispersion to 0.121 mag, the lowest scatter Cepheid \PL relation measured outside the Local Group. Such uncertainty in the dispersion for a sample this size is 0.015--0.020 mag. 
While we expect low dispersion in the absence of crowding noise, this appears even lower than expected (at 2--3$\sigma$ confidence) as compared to the spiral. It may be that the diffuse nature of NGC$\,$3447A may further improve the quality of the {\it HST} measurements in $V-I$ and as used for phase correction. For NGC$\,$3447A, our phase corrections decreased the scatter from 0.137 mag to 0.121 mag, a reduction of 0.066 mag (vs.~the spiral which dropped from 0.201 mag to 0.194 mag, a reduction of 0.049 mag). Only the Large Magellanic Cloud (LMC) and the Small Magellanic Cloud (SMC) \PL relations observed with {\it HST} in the NIR $W_H$ system exhibit lower dispersion than NGC$\,$3447A \citep{Riess:2019,Breuval:2024}. We estimate the dispersion in NGC$\,$3447A is approximately composed (in quadrature) of signal-to-noise ratio (0.03 mag), instrinsic \cite[related to the instability strip width $\leq$0.07 mag;][]{Riess:2019}, and remaining phase noise (0.10 mag). 

The apparent reddening is also low for the Cepheids in NGC$\,$3447A. Measuring $E(V-I) \approx (V-I)- (0.75 +  0.25 (\log P-1))$ (R22) or equivalently $E(F555W-F814W) \approx (F555W-F814W)- (0.85 +  0.34 (\log P-1))$, we find $E(F555W-F814W)=0.10 \pm 0.03$ mag in the spiral host NGC$\,$3447, and $E(F555W-F814W)=$ 0.02 $\pm 0.03$ in NGC$\,$3447A. The difference between the components of NGC$\,$3447  is $\Delta E(F555W-F814W) = 0.07$ mag, in good agreement with the color difference seen between the RSG branch in each component. Subtracting the MW reddening in the direction of NGC$\,$3447, $E(V-I)=0.036$ mag \citep{Schlafly:2011}, leaves negligible remaining, mean interstellar extinction for NGC$\,$3447A (likely due to efficient star formation in the interaction) and less than seen for Cepheids in the LMC, the SMC, or other SN hosts. 

\begin{deluxetable}{lllrc}[h]
\tablecolumns{5}
\tablewidth{\textwidth}
\tablecaption{Distance moduli ($\mu$), errors, sample sizes, and $P$--$L$ dispersions\\for various Cepheid sets in the system.}

\tablehead{\colhead{Dataset} & \colhead{\boldmath$\mu_0$} & \colhead{Error}& \colhead{N} & \colhead{$W_H$ $P$--$L$ $\sigma$}\\
 & \multicolumn{2}{c}{\textrm{[mag]}} & & \colhead{\textrm{[mag]}}}
\startdata
{\it HST} all (R22)           & 31.947 & $\pm$ 0.045 & 101 & 0.30 \\
{\it HST} refit$^*$, same {\it JWST} slope and color           & 31.922 & $\pm$ 0.040 & 101 & 0.30 \\
{\it JWST} N3447 All         & 31.934 & $\pm$ 0.023 & 144 & 0.174 \\
{\it JWST} N3447 Spiral      & 31.909 & $\pm$ 0.030 & 63  & 0.194 \\
{\bf {\it JWST} N3447A (``Perfect Host'') }          & 31.911 & $\pm$ 0.025 & 55  & {\bf 0.121} \\
\enddata
\tablecomments{*: {\it HST} refit uses slope =$-$3.25 following mean of {\it JWST} observations rather than $-$3.30 used by R22 for {\it HST} $F160W$,  with the goal of producing a slope-independent comparison by adopting the same slope for both telescopes. Primary impact is from  $\Delta {\rm log} P=0.4$ relative to NGC$\,$4258, $\Delta$ slope = 0.05, net $\sim 0.02$ mag. Distance errors include the measurement of Cepheids in the geometric reference, NGC 4258, $\sigma \sim 0.02$ mag.}
\end{deluxetable}

The Cepheids measured with {\it JWST} within the confines of the spiral and tidal host give similar results at $\mu=31.91 \pm 0.03$ mag, with the full sample ($N=144$) yielding $\mu=31.93 \pm 0.02$ mag. The difference between the two JWST fields is formally 0.002 $\pm 0.028$ mag.  The uncertainty here is very small because the measurement between components of NGC 3447 is purely {\it differential}, independent of either the distance to NGC 4258 or the measurement of its own Cepheids. 
 The {\it JWST} measures are also consistent with the measurement from {\it HST}  (calibrated by NGC$\,$4258, \citealt{Riess:2024b}) of $\mu=31.947 \pm 0.045$ mag, though closer by 0.03 mag, as shown in Table 1.
The difference between {\it JWST} and {\it HST} are even smaller if we refit {\it HST} Cepheid data to use the same \PL slope used for {\it JWST} of $-$3.25 (vs. R22 which used $-$3.30) and employ the color transformation between {\it HST} $F160W$ and {\it JWST} $F150W$ as by \cite{Riess:2024} which yields $\mu_{\it HST}=31.922 \pm 0.040$ mag and a difference $HST-JWST$=0.01 mag.

The SN Ia in this host, SN$\,$2012ht, yielded a luminosity calibration of $M_B=-19.20 \pm 0.10$ mag from the {\it HST} measurement by R22, about 1$\sigma$ fainter than the sample mean of 42 SNe~Ia which yields H$_0=73.0$ km/s/Mpc. As this host lies near the mean distance of the full SH0ES sample, a hypothetical, linear bias with distance modulus as required to solve the Hubble tension would have a value of $5\,\log(73/67)=0.17$ mag at the middle distance of the sample (or this constant value at any distance). NGC$\,$3447A is 3.6$\sigma$ brighter/closer and thus inconsistent with that hypothesis as shown in Figure 8.
Alternatively, SN$\,$2012ht is $0.24 \pm 0.10$ mag fainter (and in the absence of crowding) than what would solve the Hubble tension.

\subsubsection{JWST After Cycle 2}

\begin{figure}[b!] 
\begin{center}
\includegraphics[width=0.95\textwidth]{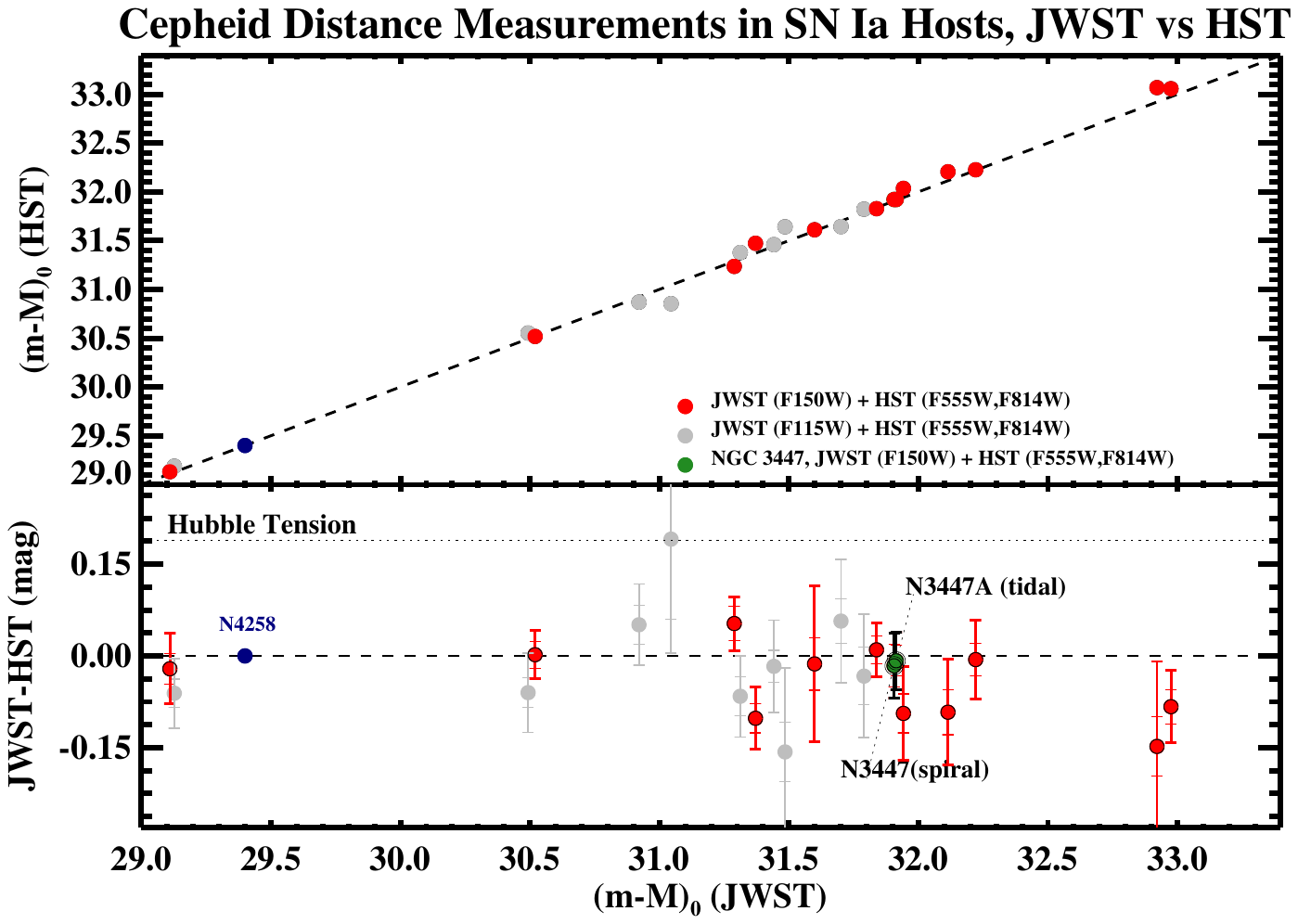}
\end{center}
\caption{Comparison of distances to 19 hosts of 24 SNe Ia measured with {\it HST} and {\it JWST} anchored by the same geometric
distance reference, NGC$\,$4258. The lower plot shows the differences in the measurements from the two telescopes. Black illustrates
the comparison for the baseline system used to measure H$_0$, $W_{H
V,I}$ where $H=F150W$ from {\it JWST} NIRCam. Red shows $W_{J
V,I}$ where $J=F115W$.  Green gives the two components of NGC$\,$3447 measured in $W_H$. The bottom plot shows a constant value of $5\,\log(73/67.5)=0.17$ mag labeled ``Hubble Tension'' that would resolve the tension (but is strongly ruled out).}
\label{fg:discomp} 
\end{figure}

For comparison we also show in Figure \ref{fg:discomp} the differences in distance, both calibrated to NGC$\,$4258, between {\it JWST} and {\it HST} for all SN~Ia hosts observed by {\it JWST} in Cycles 1 or 2.  These distance measurements in the $W_H$ filter system are included here (see Appendix) in this study of NGC$\,$3447 to ensure that all data products utilize the same {\it JWST} pipeline (Data Build 10). (Future work will present their all-band analysis.)

For hosts measured with {\it JWST} $F150W$ in the $W_H$ reddening-free system, the reference result for {\it HST} is fitted at the same slope ($-$3.25) and with the {\it HST} to {\it JWST} color transform as described by \cite{Riess:2024}. For the hosts measured with {\it JWST} $F115W$ (GO-1995, PI W. Freedman), we use a reddening-free magnitude of $W_J=J-0.64(V-I)$ (where $J=F115W$) and a slope with log(period) = $-$3.20. The {\it HST} $F160W$ and {\it JWST} $F115W$ bandpasses are too far apart to assume the same slope, so the reference result for {\it HST} comes from the global solution of R22 (calibrated to NGC$\,$4258) and as a result has a bit greater uncertainty. (One can compare {\it HST} and {\it JWST} measures of $W_H$ which assume the same slope, reducing uncertainty in the comparison.)
We find the mean difference between the distances measured here from $W_J$ and those initially presented by \cite{Freedman:2024} (version 1 only) is 0.01 mag. 

For the three most crowded hosts (new for Cycle 2), NGC$\,$5861, NGC$\,$3147, and NGC$\,$2525 at $32 < \mu < 33$~mag ($25 < D < 40$~Mpc), the {\it JWST} measurements are consistent with {\it HST} and significantly closer (brighter) than would resolve the Hubble tension. With greatly reduced crowding, the {\it JWST} results overall provide very strong evidence that the {\it HST} Cepheid photometry is reliable and not biased by crowding, with a mean difference of $-$0.022 $\pm 0.029$ mag for 19 hosts of 24 SNe~Ia, more than half the entire SH0ES {\it HST} sample from R22. The evidence against a crowding solution exceeds 8$\sigma$ and certainly exceeds the significance of the Hubble tension itself, making it an unlikely route to enlist in a resolution.
We also see no evidence of a trend between {\it HST} and {\it JWST} with distance, $-0.005 \pm 0.014$ mag per mag, nor a trend of calibrated SN~Ia $M_B$ with {\it JWST} crowding as shown in the inset of Figure \ref{fg:crowd}.

Combining the distance-ladder data presented by R22 including 42 SNe~Ia in 37 hosts with the {\it JWST} measurements of 24 SNe~Ia in 19 hosts with covariance weighting yields H$_0=73.49 \pm 0.93$ km~s$^{-1}$~Mpc$^{-1}$.  Further combining these with the {\it HST} and {\it JWST} measurements of the tip of the red giant branch (TRGB) of 35 SNe~Ia, the baseline data (CCHP calibration of $M_I$) given by \cite{Li:2025},  with covariance weighting of all common data sources yields H$_0=73.18 \pm 0.88$ km~s$^{-1}$~Mpc$^{-1}$ for 55 SN~Ia calibrators.  At this point, {\it JWST} contributes less than {\it HST} to this measurement owing to the availability of a single anchor, NGC$\,$4258, with direct observations.  Excluding {\it Gaia} parallaxes and the LMC and SMC detached eclipsing binaries leaves both {\it HST} and {\it JWST} anchored by NGC$\,$4258 and results in H$_0=72.85 \pm 1.30$ km~s$^{-1}$~Mpc$^{-1}$.  We do not include the spectroscopic SN Ia standardization results from \cite{Murakami:2023} here
as these require a more complex treatment of SN calibrator-to-SN in the Hubble flow analysis than undertaken here.

\section{Discussion}

The tidal companion to NGC$\,$3147, NGC$\,$3447A, provides a unique platform for resolving Cepheids in an SN~Ia host, one of vanishing background noise. Because measurements of Cepheids at large distances are generally limited by crowding noise before faintness, it is plausible to consider extending Cepheid measurements in a young tidal environment like NGC$\,$3447A to distances approaching $\sim 100$ Mpc, thus without the need to calibrate a hosted SN~Ia and reducing the distance ladder by a rung \citep{2025arXiv250702056A}. 

Were it not for the Hubble tension, the topic of Cepheid crowding would rightly receive little attention. This is because crowding has long been a ``solved problem'' in terms of systematic error, thanks to the advent of digital photometry and the use of crowded-field (multi-PSF) fitting algorithms \citep{Ferrarese:2000}. Fundamentally, there is nothing especially complex about Cepheid crowding. The Cepheids we observe appear superimposed on a crowded background of stars, with the degree of crowding determined by our line of sight to the host galaxy. This setup is well described by random superposition, and thus the process can be effectively mimicked and corrected by the addition and measurement of artificial stars. Like many problems in astronomy, Cepheid crowding is ultimately a background estimation problem -- one that can be fully simulated without unknown parameters\footnote{Accurate Cepheid photometry can be obtained using a small aperture centered on the Cepheid with an annulus to estimate the {\it mean} sky background. However, such measurements lack precision because they do not separate blended sources or weight the flux using the PSF. Multi-PSF fitting offers greater precision but introduces a bias: the sky level defined after source removal is lower than the true mean, which should include fully superimposed blended sources. The combination of multi-PSF fitting and artificial-star-based debiasing provides a route to the most precise and accurate Cepheid photometry currently available.}.
It is therefore unsurprising that Cepheid photometry from {\it HST} matches that from {\it JWST}, where crowding is minimal, as shown here and previously \citep[R23;][]{Riess:2024}.

Even before {\it JWST}, comparisons of extragalactic Cepheid amplitudes with those observed in the MW demonstrated that observed and corrected crowding fully accounted for the amplitude differences, leaving little room for error \cite{Riess:2020,Sharon:2024}. Additionally, the close agreement between Cepheid-based distances and those derived from TRGB, JAGB, and Mira variables -- all calibrated to the same geometric anchor, NGC$\,$4258 -- supports the conclusion that these distance measures are free of systematics at the 1\% level \citep{Riess:2024b,Freedman:2024}. Farther out, replacing SNe~Ia with alternative far-field distance indicators (surface brightness fluctuations, masers, fundamental plane, Tully-Fisher, SNe~II) in the distance ladder generally increases the inferred value of H$_0$ \citep{2023ApJ...944...94T, Scolnic:2025,2025ApJ...987...87J}. Consequently, attributing the Hubble tension to systematic errors in measurement has become increasingly untenable, making a more fundamental cause the leading possibility.

\section{Acknowledgments \label{sec:acknowledgments}}

We are indebted to all who spent years and even decades bringing {\it JWST} to fruition. This research has made use of NASA’s Astrophysics Data System. D.S. is supported by Department of Energy grant DE-SC0010007, the David and Lucile Packard Foundation, the Templeton Foundation, and the Sloan Foundation. G.S.A. acknowledges financial support from {\it JWST} grant GO-2875. C.D.H. acknowledges financial support from {\it HST} grants GO-16744 and GO-17312.
A.V.F.'s research group at U.C. Berkeley acknowledges financial assistance from the Christopher R. Redlich Fund, as well as donations from William Draper, Timothy and Melissa Draper, Briggs and Kathleen Wood, Sanford Robertson (T.G.B. is a Draper-Wood-Robertson Specialist in Astronomy), and numerous other donors. 

Some of the data presented herein were obtained at the W. M. Keck Observatory, which is operated as a scientific partnership among the California Institute of Technology, the University of California, and NASA; the observatory was made possible by the generous financial support of the W. M. Keck Foundation. We thank WeiKang Zheng for assistance with the Keck/LRIS observations.

If something does not make sense, or you think may be in error, please email ariess@stsci.edu with questions.

\clearpage

\bibliographystyle{aasjournal}
\bibliography{bibsh0es}

\appendix

\begin{figure}[b!] 
\begin{center}
\includegraphics[width=\textwidth]{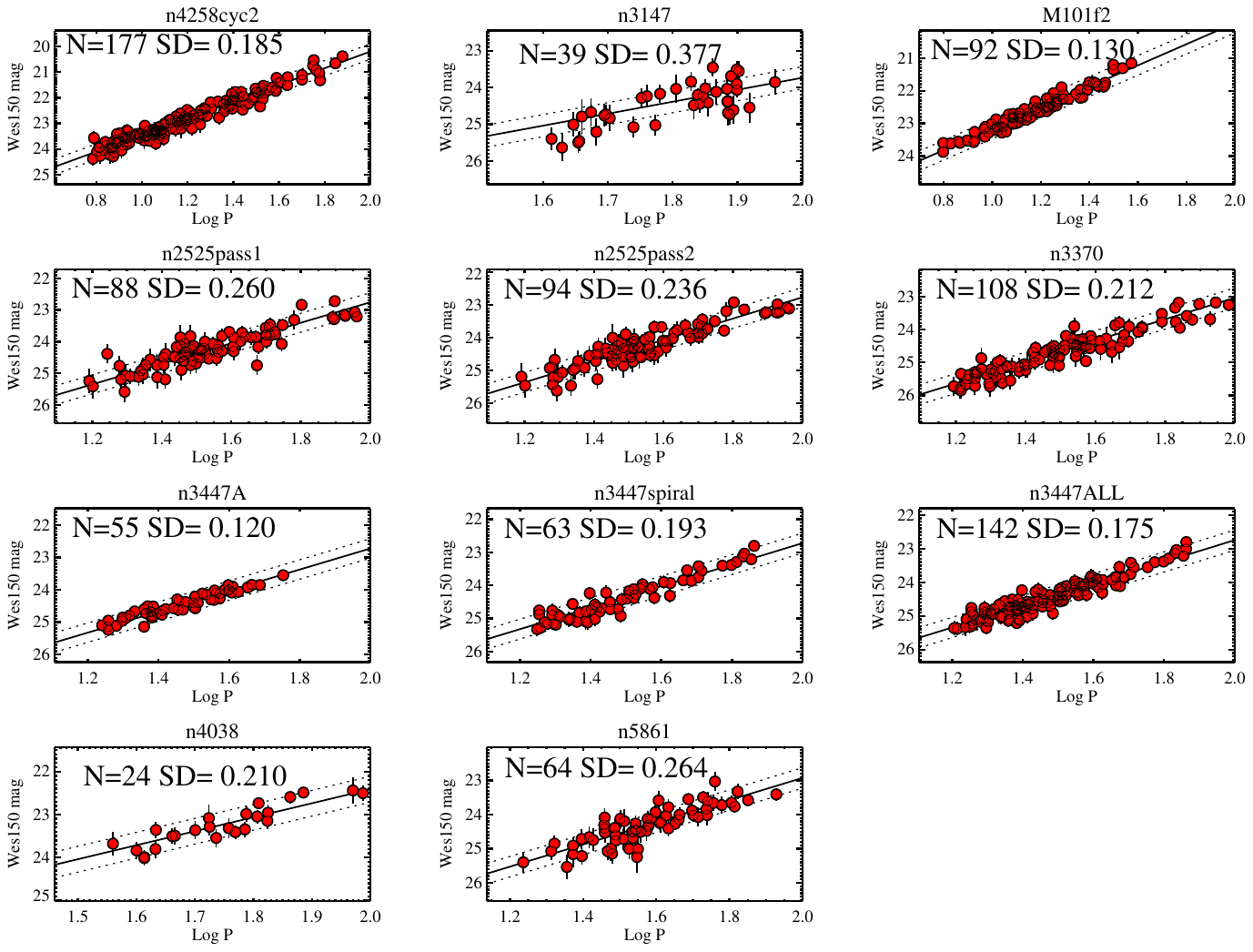}
\end{center}
\caption{Wesenheit relations for $F150W$ for Figure 9.  Additional hosts and color combination relations may be found in \cite{Riess:2024}.  $F150W$ Data is from GO programs 1685 and 2875 (PI Riess) with the exception of NGC 4038 and M101 and all $F555W-F814$ colors are from the HST SH0ES project.  All data is phase-corrected using an additional {\it JWST} short-wavelength color as described in section 3.2.1}
\label{fg:wes150} 
\end{figure}

\begin{figure}[b!] 
\begin{center}
\includegraphics[width=\textwidth]{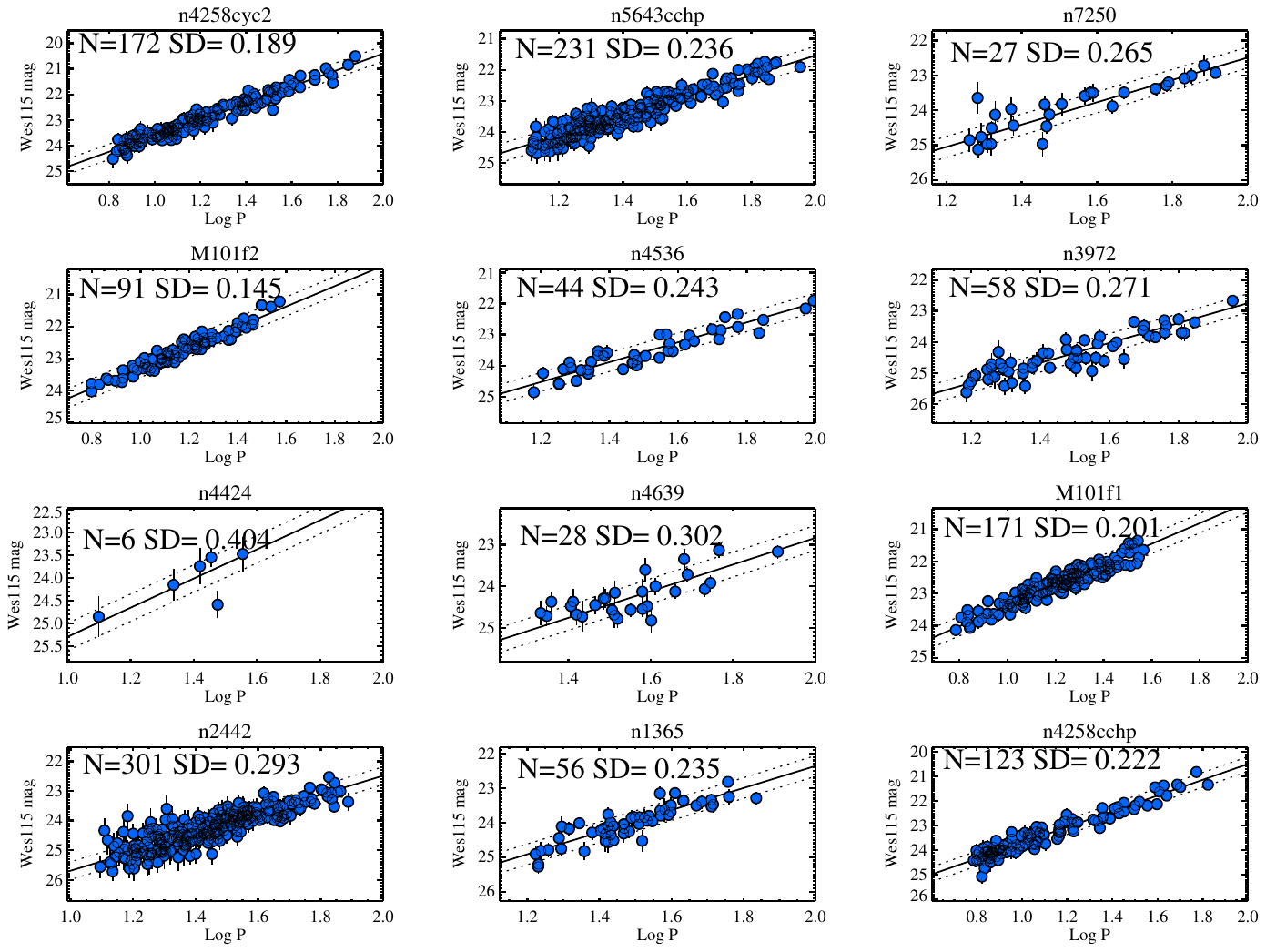}
\end{center}
\caption{Wesenheit relations for $F115W$ Wesenheit relations for $F115W$ for Figure 9.   $F115W$ data is from GO program 1995 (PI Freedman) with the exception of ``NGC 4258cyc2'' (GO 2875, PI Riess) and M101f2 (PI. Huang), and all $F555W-F814$ colors are from the HST SH0ES project.}
\label{fg:wes115} 
\end{figure}

Here we provide the measurements of Cepheids whose summary results are shown in Figure \ref{fg:discomp}. Table A provides the distance measures and the Cepheid Wesenheit \PL are shown in Figure \ref{fg:wes150} for $W_H$ magnitude measurements and in Figure \ref{fg:wes150} for $W_H$ magnitude measurements and in Figure \ref{fg:wes115} for $W_J$ magnitude measurements.  Table B provides the Cepheid photometry for the Cepheids in NGC 3447.  The measurements in $W_H$ for the same target tend to have less scatter than those in $W_J$ by $\sim$ 30\%, a consequence of the additional JWST color, usually $F090W$ obtained by the SH0ES program (2875) (but not in the CCHP program, GO 1995) which is used to improve knowledge of the Cepheid phase as described in $\S$ 3.2.1.   The one example where short-wavelength color information is available for $W_J$ and $W_H$ for the same target, ``NGC4258cyc2'' the dispersions are highly similar and both are smaller than ``n4258cchp'' by 20\%.  Likewise for the M101 field where the color information reduces the scatter by 40\%.

Six Additional $W_H$ \PL are presented in \cite{Riess:2024} for NGC 5643, 1448, 1559, 5468, 5584 and 4258 which employ two {\it JWST} epochs.

\section{Data tables}

\begin{table}[ht]
\centering
\begin{tabular}{lccccc}
\hline
Host & {\it JWST} & $\sigma$ & {\it HST} & $\sigma$ & JWST\_Filter \\
\hline
NGC$\,$3147     & 32.92 & 0.05 & 33.07 & 0.13 & F150W \\
NGC$\,$2525 & 31.94 & 0.03 & 32.04 & 0.07 & F150W \\
NGC$\,$5861     & 32.11 & 0.04 & 32.21 & 0.08 & F150W \\
NGC$\,$3370     & 32.22 & 0.03 & 32.23 & 0.06 & F150W \\
\hline
NGC$\,$5643 & 30.49 & 0.02 & 30.55 & 0.06 & F115W \\
NGC$\,$7250     & 31.49 & 0.05 & 31.64 & 0.13 & F115W \\
NGC$\,$4536     & 30.92 & 0.03 & 30.87 & 0.06 & F115W \\
NGC$\,$3972     & 31.70 & 0.04 & 31.64 & 0.09 & F115W \\
NGC$\,$4424     & 31.05 & 0.13 & 30.85 & 0.13 & F115W \\
NGC$\,$4639     & 31.79 & 0.05 & 31.82 & 0.09 & F115W \\
M101   & 29.13 & 0.02 & 29.19 & 0.05 & F115W \\
NGC$\,$2442     & 31.44 & 0.03 & 31.46 & 0.07 & F115W \\
NGC$\,$1365     & 31.31 & 0.03 & 31.38 & 0.06 & F115W \\
\hline
\end{tabular}
\caption{Distance moduli and uncertainties from {\it JWST} and {\it HST} for various hosts and filters.}
\label{tab:moduli}
\end{table}

\startlongtable
\begin{deluxetable*}{lrrrrrrrr}
\tabletypesize{\scriptsize}
\tablewidth{0pt}
\tablenum{B}
\tablecaption{Photometric Data for Cepheids in NGC 3447\label{tb:phot}}
\tablehead{\multicolumn{1}{c}{Host} &  \multicolumn{1}{c}{ID} & \multicolumn{1}{c}{RA} & \multicolumn{1}{c}{Dec} & \multicolumn{1}{c}{log P} & \multicolumn{1}{c}{{\it F150W}} & \multicolumn{1}{c}{$\sigma$}  & \multicolumn{1}{c}{{\it V}$-${\it I}$\,^a$} & \multicolumn{1}{c}{$\sigma$}}
\startdata
N3447 &        1055 &  163.348250  & 16.79841  &  1.4119  &  25.155  &  0.111   &  1.279  &  0.19   \\
\hline
\enddata
\tablecomments{$a$: {\it F555W}$-${\it F814W}.  We note the provided magnitudes from JWST are phase corrected.  This table will be available in its entirety upon publication in digital form.}
\phantom{}
\vspace{-48pt}
\end{deluxetable*}

\end{document}